\documentclass[aps,pra,showpacs,twocolumn,superscriptaddress]{revtex4-1}
\usepackage{graphicx}% Include figure files
\usepackage{dcolumn}% Align table columns on decimal point
\usepackage{bm}% bold math

\usepackage[alsoload=hep]{siunitx}	
\usepackage[colorlinks=true, citecolor=blue,allcolors=blue]{hyperref}

\newunit{\atomicunit}{\textmd{a{.}u{.}}}			% define a.u.

\begin{document}
\title{Circular Dichroism in Atomic Resonance-Enhanced Few-Photon Ionization} 

\author{A.H.N.C.\ De Silva}
\affiliation{Physics Department and LAMOR, Missouri University of Science \& Technology, Rolla, MO 65409, USA}

\author{T.\ Moon}
\affiliation{Department of Physics, Kennesaw State University, Kennesaw, Georgia 30144, USA}

\author{K.L.\ Romans}
\affiliation{Physics Department and LAMOR, Missouri University of Science \& Technology, Rolla, MO 65409, USA}

\author{B.P.\ Acharya}
\affiliation{Physics Department and LAMOR, Missouri University of Science \& Technology, Rolla, MO 65409, USA}

\author{S.\ Dubey}
\affiliation{Physics Department and LAMOR, Missouri University of Science \& Technology, Rolla, MO 65409, USA}

\author{ K.\ Foster}
\affiliation{Physics Department and LAMOR, Missouri University of Science \& Technology, Rolla, MO 65409, USA}

\author{O.\ Russ}
\affiliation{Physics Department and LAMOR, Missouri University of Science \& Technology, Rolla, MO 65409, USA}

\author{C.\ Rischbieter}
\affiliation{Physics Department and LAMOR, Missouri University of Science \& Technology, Rolla, MO 65409, USA}

\author{N.\ Douguet}
\affiliation{Department of Physics, Kennesaw State University, Kennesaw, Georgia 30144, USA}

\author{K.\ Bartschat}
\affiliation{Department of Physics  and Astronomy, Drake University, Des Moines, Iowa 50311, USA}

\author{D.\ Fischer}
\affiliation{Physics Department and LAMOR, Missouri University of Science \& Technology, Rolla, MO 65409, USA}

%\author{many others}
%\affiliation{Physics Department and LAMOR, Missouri University of Science \& Technology, Rolla, MO 65409, USA}

%\author{even more}
%\affiliation{Department of Physics  and Astronomy, Drake University, Des Moines, Iowa 50311, USA}

\date{\today}% It is always \today, today,

\begin{abstract}
We investigate few-photon ionization of lithium atoms prepared in the polarized 2$p\,$($m_\ell=\!+1$) state
when subjected to femtosecond light pulses with left- or right-handed circular polarization 
at wavelengths between 665\,nm and 920\,nm. We consider whether ionization proceeds 
more favorably for the electric field co- or counter-rotating with the initial electronic current density.
Strong asymmetries are found and quantitatively analyzed in terms of ``circular dichroism" ($CD$). 
 While the intensity dependence of the measured $CD$ values is rather weak throughout the investigated regime, 
 a very strong sensitivity on the center wavelength of the incoming radiation is observed. 
 While the co-rotating situation overall prevails, the counter-rotating geometry is strongly 
 favored around 800\,nm due to the 2$p$-3$s$ resonant transition, which can only be driven by counter-rotating fields.  
 The observed features provide insights into the helicity dependence of light-atom 
 interactions, and on the possible control of electron emission in atomic few-photon ionization 
by polarization-selective resonance enhancement.
\end{abstract}

%\pacs{34.50.Fa}

\maketitle

\section{Introduction}
The response of matter to circularly polarized light can depend on the helicity of the light's polarization. 
This phenomenon is referred to as circular dichroism ($CD$) and is widely exploited, e.g., in the analysis of chiral molecules 
(see Ref.~\cite{Berova2014} for an overview). 
Biomolecules, such as sugar or amino acids, are prominent examples of chiral targets  for which single- 
\cite{Boewering2001, Garcia2013, Nahon2015} and multi-photon ionization \cite{Goetz2019, Lux2012, Lux2015},
as well as tunnel ionization in intense laser fields \cite{Beaulieu2018}, exhibit a dependence
on the light's helicity even for randomly oriented molecules. These dichroic differences have even been 
suggested to be the symmetry-breaking cause of life's homochirality \cite{Jorissen2002,Tia2013}. Nevertheless, these effects are 
typically weak and vanish completely in the electric dipole approximation after integration over all possible target orientations. 

For spatially aligned systems, in contrast, a $CD$ signal occurs already for electric dipole transitions, 
resulting in much stronger asymmetries \cite{Tia2017}. The most fundamental realizations of
such oriented targets are polarized atoms with a magnetic quantum number $m\neq 0$.
In this situation, the helicity dependence is a result of either equal or opposite rotations of the electric field 
with respect to the initial current density of the active electron. Due to their fundamental importance,
these benchmark systems have recently attracted considerable 
interest~\cite{Herath2012,Mancuso2017,Eckart2018,Sainadh2019,Mazza2014,Ilchen2017,GrumGrzhimailo2019,Silva2021}
in the study of the interaction between handed light and handed matter.
 
Atomic circular dichroism is quantitatively defined as
\begin{equation} 
CD=\frac{P_+-P_-}{P_++P_-},
\label{eq:Cd}
\end{equation}
where $P_+ $ and $P_-$ are the cross sections for the co- and counter-rotating  light polarizations respectively. 
In previous studies, it was found that there is no unambiguous answer to the fundamental question whether 
ionization proceeds more likely for the co- or the counter-rotating case. In the simplest case of one-photon ionization by a 
weak field, this question was already answered by 
Bethe and Salpeter, whose theoretical framework \cite{Bethe77} delivers a $CD$ value of close to 1, 
meaning that the co-rotating geometry is strongly favored \cite{Barth2011,Herath2012}. 
To the contrary, in the non\-adiabatic tunneling regime, i.e., for intense fields, the ionization 
cross section was found to be  larger for a counter-rotating field, thus leading
to negative $CD$ values. 

In multi-photon ionization, the sign of the $CD$ value features a 
dependence on the intensity of the ionizing field \cite{Ilchen2017,Bauer2014}. 
This intensity dependence can be rather strong \cite{GrumGrzhimailo2019} due to ``Freeman'' 
resonances \cite{Freeman1987}, where transient resonance enhancement occurs due to AC Stark shifts of inter\-mediate levels.

In a recent Letter \cite{Silva2021}, two-photon ionization of polarized atoms was 
investigated in the case where resonance enhancement did not significantly affect the ionization rates. 
Lithium atoms were prepared in the 2$p\,$($\left| m\right|=1$) state and then ionized by a field (671\,nm) tuned to the  
$2s-2p$ resonance transition. A positive and weakly intensity-dependent value of the CD was found 
throughout the investigated intensity regime, while strong helicity-selective Autler-Townes 
shifts were observed for the co-rotating geometry due to the coupling of the excited initial target 
state with its ground state. This peculiar feature allowed us to control not only the ionization 
probability, but also the photo\-electron energy via the field polarization and intensity.

In the present study, we report an extension of our earlier work \cite{Silva2021}, 
now altering not only the intensity, but also the wavelength of the ionizing field. Specifically, 
we scan a wavelength range around the $2p$-$3s$ transition (at about 812\,nm). Close to this resonance,
the 2$p$ state is ionized by the absorption of three photons, and resonance 
enhancement through the $3s$ state, as well as Autler-Townes shifts due to the $2p$-$3s$ coupling, 
are expected to occur only for the counter-rotating geometry. 
In combination with the results reported in \cite{Silva2021}, the present work aims to 
understand the role of resonance enhancement in circular dichroism and the sensitivity 
of dichroic asymmetries to wavelength change of the ionizing radiation. 

The experimental results are compared with  
theoretical predictions obtained by solving the time-dependent Schr{\"o}dinger equation (TDSE). These results exhibit overall
good agreement with the measured $CD$ values. As in our earlier study, 
the present system represents a fundamental realization of a chiral system and, 
therefore, provides fundamental insights into implications of asymmetries in light-matter interaction.

%Hence, we investigated polarized lithium atoms at an ultra-cold temperature with a single electron in the valence shell. There have been several recent studies on this symmetry breaking phenomenon with resonance-enhanced multi-photon ionization(REMPI)\cite{Grum} and with non-resonance enhanced multi-photon ionization \cite{De Silva} with no electrons being generated near threshold. This paper also focuses on a study of non-resonance enhanced multi-photon ionization without threshold electrons to make the data analysis easier and the circular dichroism clean without contamination due to resonances. The experiment was carried out anticipating a negative CD value in contrast to our previously reported experiment due to the lower photon energy at 790nm hitting the 3s resonance state only in the counter-rotating case, which was assumed to have a higher cross-section than the corotating case for the ionization from the 2$p$ excited state.    

\section{Experimental Setup}
Our experiment was performed with the same apparatus described in \cite{Silva2021}. 
It consists of three components: An all-optical lithium atom trap (AOT) \cite{Sharma2018}, 
a tunable femtosecond light source, and a reaction microscope that allows us to measure the 
electron and recoil ion momenta after an ionization process \cite{Hubele2015, Thini2020}. 

In the AOT, lithium atoms are laser-cooled to a temperature of about 1\,mK in the field of 
continuous-wave laser beams near the 2$s$-2$p$ resonance at 671\,nm and trapped in a volume 
of approximately 1\,mm diameter. In steady state, about 25\,\% of the atoms in the 
trap are in the excited 2$p$ state. A fraction of about 93\,\% of these atoms are in a single magnetic sub\-level 
with $\left|m_\ell\right|=1$ with respect to the direction of a weak homogeneous 
magnetic field (about 5\,Gauss), which is referred to as the $z$ direction below. 
As detailed in \cite{Sharma2018}, this high degree of polarization is achieved by an appropriate choice of the frequency as well as
the polarizations of the cooling laser beams exploiting the Zeeman splitting of the magnetic
sub\-levels as well as the dipole selection rules, respectively. This way an optical-pumping 
scheme is employed resulting in an average increase of the magnetic quantum number for each absorbed photon. 

\begin{figure}
\centering
\includegraphics[width=\linewidth]{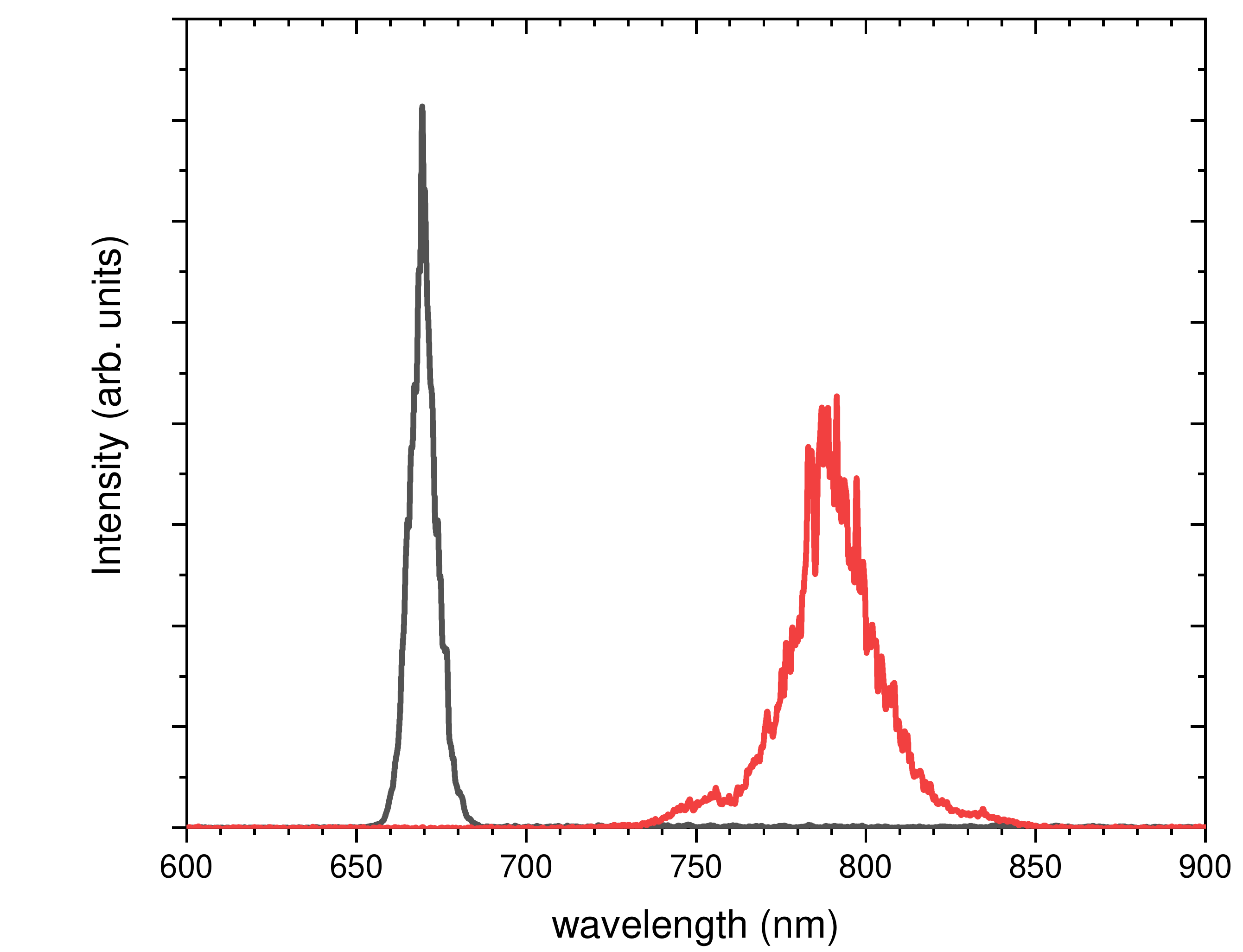}
\caption{Typical emission spectra of the femtosecond light source with center 
wavelengths tuned to 665\,nm (black curve) and 795\,nm (red curve).  \label{fig:Spectrum}}
\end{figure}

The femtosecond laser is based on a Ti:Sa oscillator with two non\-collinear 
optical parametric amplifier (NOPA) stages similar to the setup reported in \cite{Harth2017}. 
The oscillator provides pulses with wavelengths ranging from 660\,nm to about 1000\,nm with a pulse 
duration of 5\,fs, a repetition rate of 80\,MHz, and an average power of 200\,mW. 
These pulses are stretched by passing through a combination of sapphire and fused 
silica windows to a duration in the order of about 1\,ps. These strongly chirped 
pulses are superposed with a synchronized pump beam in two NOPA stages. The pump 
beam operates at a wavelength of 515\,nm with a rate and duration of 200\,kHz and 
250\,fs, respectively. Changing the relative timing between the pump pulse and 
the chirped oscillator pulse allows us to control the part of the spectrum that is 
amplified in the NOPA stages. In this way, the center wavelength can be tuned over
the full wavelength range of the oscillator. Behind the NOPA stages, the pulses are 
 re\-compressed by a set of chirped mirrors to a pulse length of several tens of~fs. 
 The laser is guided through the vacuum chamber at a small angle (10$^\circ$) with 
 respect to the $z$ direction and focused on the target cloud with a beam waist 
 of 50\,$\mu$m. Peak laser intensities of up to $I = 1.3\times 10^{12}$\,W/cm$^2$ 
 were reached. Two typical spectra for center wavelengths of 670\,nm and 790\,nm 
 are shown in Fig.~\ref{fig:Spectrum}.

%Due to an  target at a temperature of of 1mK, a highly intense circularly polarized femtosecond laser pulse of duration down to 65fs, a center wavelength of 790nm, and a bandwidth of 10nm. The lithium target is prepared in an all-optical trap(AOT) and momentum analyzed for the emitted electrons with an electron-ion momentum spectrometer with electron momentum resolution of 0.01\,a.u. earlier reported in \cite{Sharma}, \cite{Thini},\cite{Hubele}.  except for the change in wavelength. The atoms in the trap are excited to the 2$p$ state with a fraction of excitation of 25\% in the magnetic sub-level quantum number 1s$^2$2$p$($m_l=+1$) with a degree of polarization of 93\%.
The electron and ion momentum spectrometers are described in more detail in \cite{Hubele2015}. 
Briefly, electrons and target ions are extracted in opposite directions along the $z$ axis 
by weak homogeneous electric and magnetic fields and detected with position- and time-sensitive 
micro\-channel plate detectors. The particles' positions and time-of-flights 
allow us to determine their initial momenta \cite{Ullrich2003,Fischer2019}. As mentioned above, 
the target cloud contains atoms in the 2$s$ ground state and  the excited 2$p$ state. 
In order to separate the cross sections for the two initial states, the cooling 
lasers were switched off for 1\,$\mu$s during every other femtosecond laser pulse. 
On the one hand, this switch-off duration is much longer than the lifetime of the excited 2$p$ state (27\,ns). 
Therefore, the spectra measured for the cooling lasers being switched off correspond to 
pure ground-state (2$s$) ionization. On the other hand, the switch-off duration is much shorter than typical
timescales of the thermal motion ($v\sim 1$\,$\mu$m/$\mu$s) of the atoms in the AOT. Therefore, the number density of the target atoms 
does not change significantly during the switching cycle, and the cross sections for the 2$p$ initial state can be calculated 
by subtracting the 2$s$ cross sections from the spectra measured for the cooling lasers being 
switched on by using an appropriate scaling factor that only depends on the previously determined 
fraction of atoms being in the excited state. Notably, knowing the excited population fraction allows also
to cross-normalize the cross sections for the two initial states.

\begin{figure}
\centering
\includegraphics[width=1\linewidth]{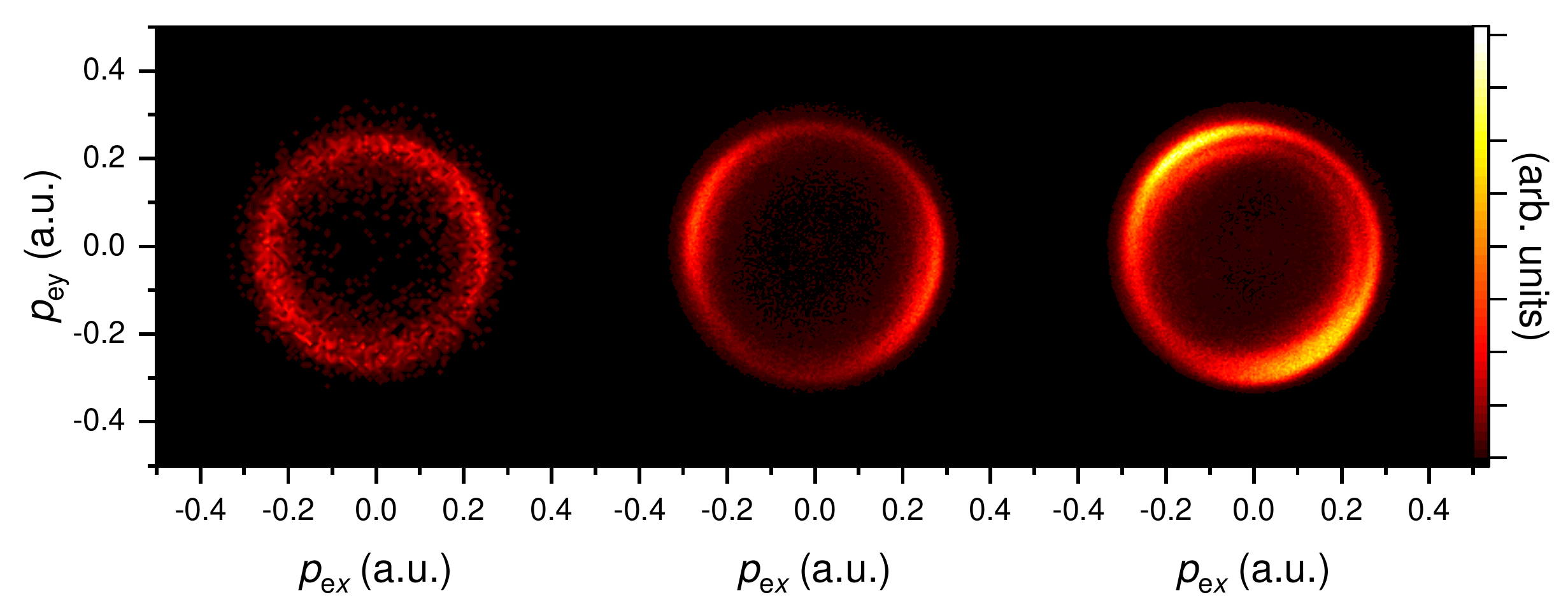}
\caption{Electron momentum distributions in the $xy$ plane for ionization of the 2$s$ (left), 
counter-rotating 2$p$ (center), and co-rotating 2$p$ state (right) at a center wavelength of 795\,nm 
and a peak intensity of 0.8$\times 10^{12}$\,W/cm$^2$. The color scale is linear, and the spectra are cross-normalized but
multiplied by a factor of 100 for 2$s$ ionization. \label{fig:momdist}}
\end{figure}

Typical electron momentum distributions integrated in the $z$ direction are shown in 
Fig.~\ref{fig:momdist} for 2$s$ ionization and 2$p$ ionization with co- and counter-rotating helicities. 
In the present experiment, a resolution of 0.01\,a.u.\ in the direction perpendicular to the femtosecond 
laser beam and of 0.005\,a.u.\ in the direction longitudinal to it was achieved. For ionization 
with light of circular polarization in the $xy$ plane, the momentum distributions shown in 
the figure should feature circular symmetry. However,  in the experiment an enhancement in 
two opposite directions is observed.  This is caused by the imperfect polarization of both 
the target atoms as well as the laser radiation, which is estimated to have an ellipticity of about 20\,\%.

\section{Theoretical model}
The experimental efforts are supported by {\it ab initio} calculations based on solving the 
time-dependent Schr\"{o}dinger equation (TDSE) in the single-active electron (SAE) picture 
for a $(n\ell)$ active electron in a He-like $1s^2$ ionic core.
Details of the potential, which is based 
on the static Hartree potential~\cite{Albright1993,Schuricke11} supplemented by phenomenological terms,
are given in \cite{Silva2021}.  We achieved an accuracy of better than 1~meV for the $n=2$ and $n=3$ ionization potentials (IPs). 
The initial state of the TDSE is propagated in the velocity gauge, and the ionization probability is computed by 
projecting the time-dependent wavefunction at the end of the pulse
onto scattering states. The earlier TDSE results exhibited excellent agreement with our experimental 
data for the photo\-electron spectra and momenta 
measured under similar experimental conditions~\cite{Silva2021}.

\section{Results and discussion}
\begin{figure}
\centering
\includegraphics[width=1\linewidth]{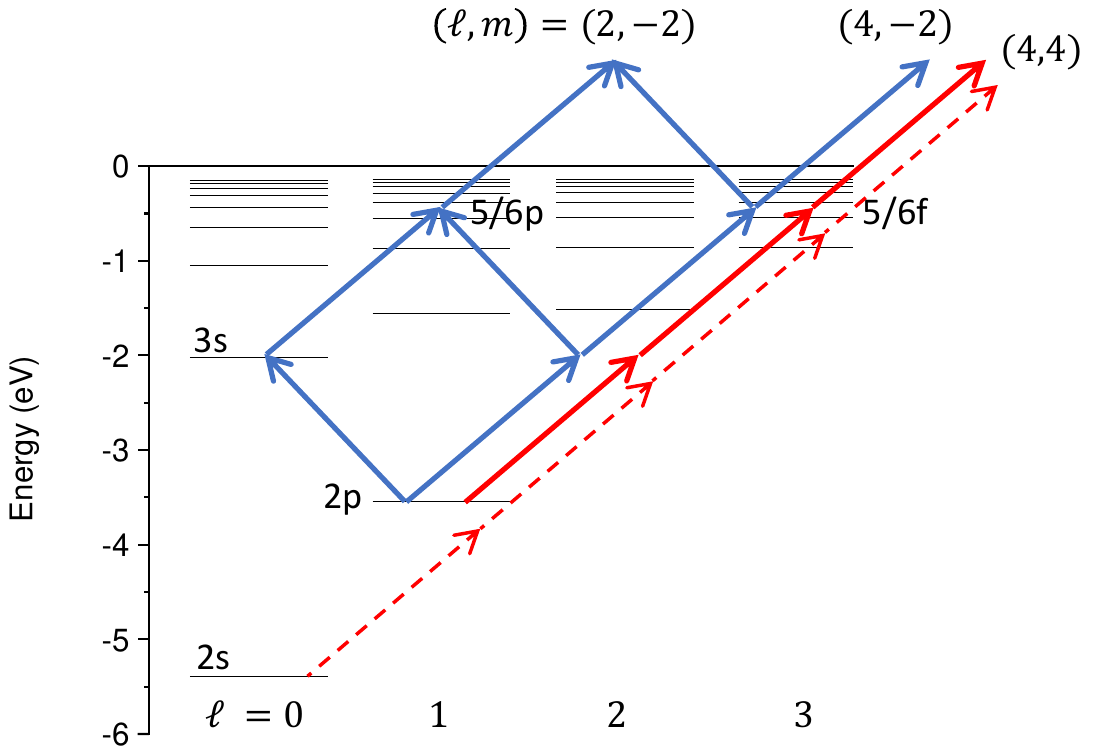}
\caption{Ionization scheme for three-photon ionization of the co- (solid red arrows) 
and counter-rotating (solid blue arrows) 2$p$ excited state as well as for four-photon 
ionization of the 2$s$ ground state (dashed red arrows). \label{fig:IonizationScheme}}
\end{figure}

For the laser wavelengths and intensities investigated in this study, 
the Keldysh parameter $\gamma$ for the valence ionization of lithium is always 
much larger than unity, such that the process can be described in the multi-photon picture
in lowest-order perturbation theory (LOPT) \cite{Lambropoulos1976}. 
Here, the absorption of one photon of right-handed circular polarization propagating 
along the quantization direction (i.e.\ the $z$ direction) results in a change of 
the magnetic quantum number $\Delta m=+1$, while the absorption of a left-handed 
circularly polarized photon leads to $\Delta m=-1$. The resulting pathways for the three-photon 
ionization of Li~(2$p\,$, $m=+1$) by co- and counter-rotating fields ($\lambda=795$\,nm) 
are depicted in Fig.~\ref{fig:IonizationScheme} as red and blue arrows, respectively. 
There is only a single possible pathway for co-rotating ionization with the electron 
being promoted to a final continuum state with $(\ell,m)=(4,4)$. For the 
counter-rotating case, in contrast, the final state is a superposition of $(\ell,m)=(2,-2)$ and $(4,-2)$ partial waves, respectively. 
Around a wavelength of 812\,nm 
(neglecting the dressing of the states in the field), resonance enhancement is expected for 
counter-rotating polarization due to the presence of the 3$s$ state. Two-photon resonances contribute, for 
either polarization, via the $n=5$ levels at about 826\,nm-829\,nm, $n=6$ at about 780\,nm, 
 $n=7$ at 760\,nm, $n=8$ at 744\,nm, etc.. The two-photon ionization threshold is 
reached at 700\,nm. For 4-photon ionization of Li(2$s$) (dashed red arrows in 
Fig.~\ref{fig:IonizationScheme}), the final angular-momentum state is, respectively, $(\ell,m)=(4,-4)$ or $(4,4)$ 
for left or right-handed circular polarizations, resulting in identical 
cross sections for the two light helicities.

\begin{figure}
\centering
\includegraphics[width=1\linewidth]{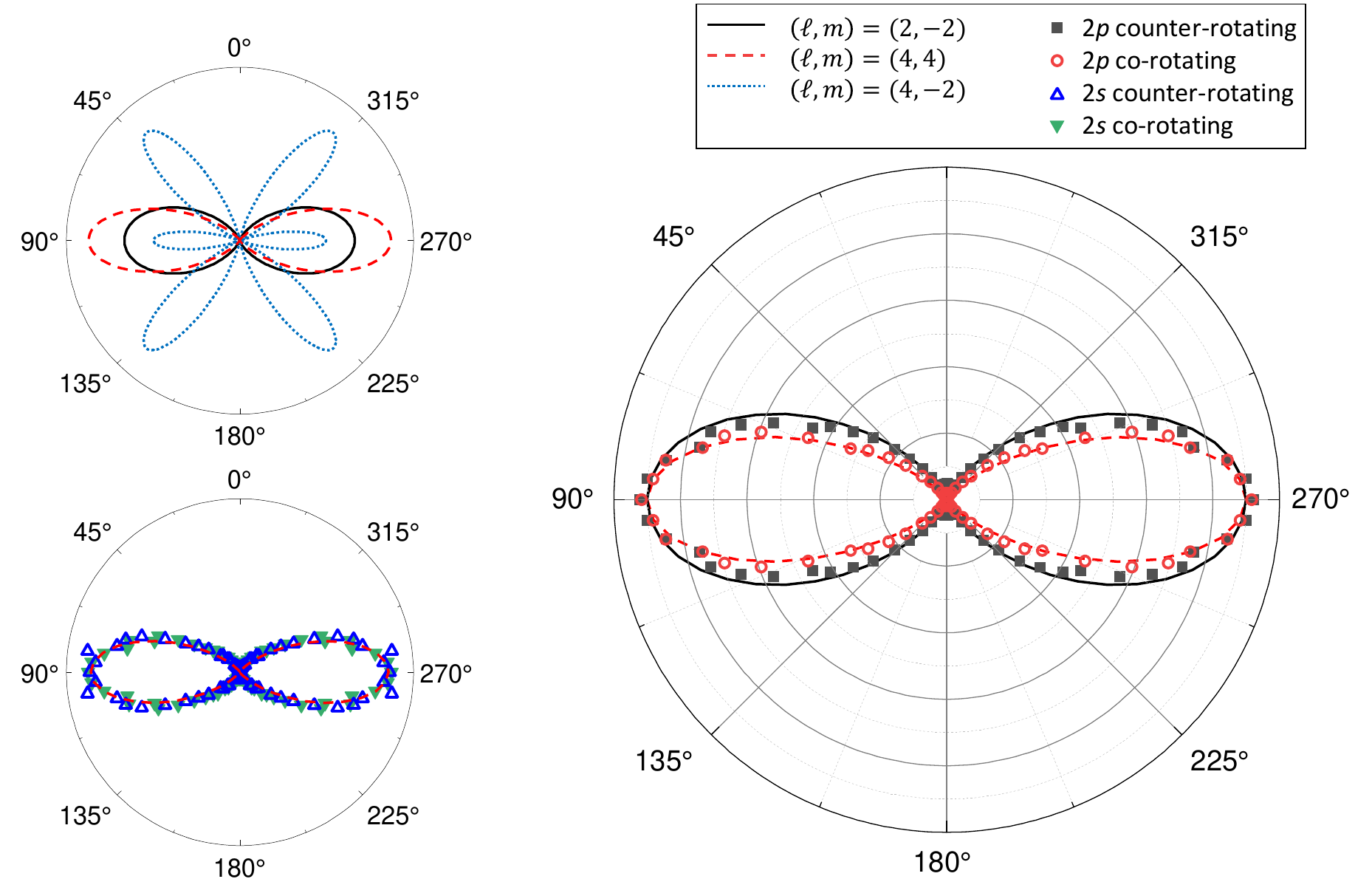}
\caption{Photoelectron angular distributions for ionization by a laser field 
with a center wavelength of 795\,nm and a peak intensity of 0.8$\times 10^{12}$\,W/cm$^2$. 
Top left: $\vartheta$ distribution of partial waves corresponding to the spherical 
harmonics $Y_4^4$ (dashed red), $Y_4^{-2}$ (dotted blue), and $Y_2^{-2}$  (solid black). 
Bottom left: Measured $\vartheta$ distribution for 2$s$ ionization for 
left- (blue open triangles) and right-handed (green solid triangles) circularly
 polarized light along with the distribution of the partial wave with $(\ell,m)=(4,4)$ 
 (dashed red line). Right: Measured $\vartheta$ distribution for 2$p$ ionization 
 for co- (black solid squares) and counter-rotating (red open circles) cases with 
 the partial waves $(\ell,m)=(4,4)$ (dashed red line) and $(2,-2)$ (solid black line).  \label{fig:PAD}}
\end{figure}

Different angular momenta contribute to the photoelectron angular 
distributions (PADs). The expected intensity distributions for the 
individual partial waves $(\ell,m)=(4,4)$, $(2,-2)$, and $(4,-2)$ are shown in the top left panel of
 Fig.~\ref{fig:PAD} as a function of the polar angle $\vartheta$. 
For $(\ell,m)=(4,4)$, the $\vartheta$ dependence is given by a sin$^8(\vartheta)$-function 
with a narrow peak perpendicular to the quantization axis. A similar behavior is 
observed for $(\ell,m)=(2,-2)$ with a slightly broader peak following a sin$^4$($\vartheta$)-distribution. 
The $g$~state with $m=-2$, in contrast, features  intensity not only at 90$^\circ$ but 
also at 40$^\circ$ and 140$^\circ$ with respect to the laser beam direction.

The comparisons to the experimental data for a center wavelength of 795\,nm are shown 
in the two other panels of Fig.~\ref{fig:PAD}. As expected, the experimental electron 
angular distributions for 2$s$ ionization are very well described by the sin$^8$-distribution 
of a $g$~state (bottom left in the figure) for both helicities. The same behavior is 
observed for 2$p$ ionization in the co-rotating geometry (Fig.~\ref{fig:PAD}, right). 
For the counter-rotating scheme, the $\vartheta$ dependence matches well with the 
sin$^4$-distribution of the $d$~state. Notably, the cross section is rather small 
at angles where the $(\ell,m)=(4,-2)$ partial wave has its highest intensities, 
indicating that the final state is vastly dominated by the $\ell=2$ wave. 
Qualitatively, the dominance of this $d$~state can be explained by ionization 
enhancement through the 2$p$-3$s$ resonance, because in LOPT this pathway necessarily results 
in final $\ell=2$ states. 
At other wavelengths, and in absence of the  2$p$-3$s$ resonance, the angular distribution 
can be vastly different (see e.g.~\cite{Silva2021}). 

\begin{figure}
\centering
\includegraphics[width=1\linewidth]{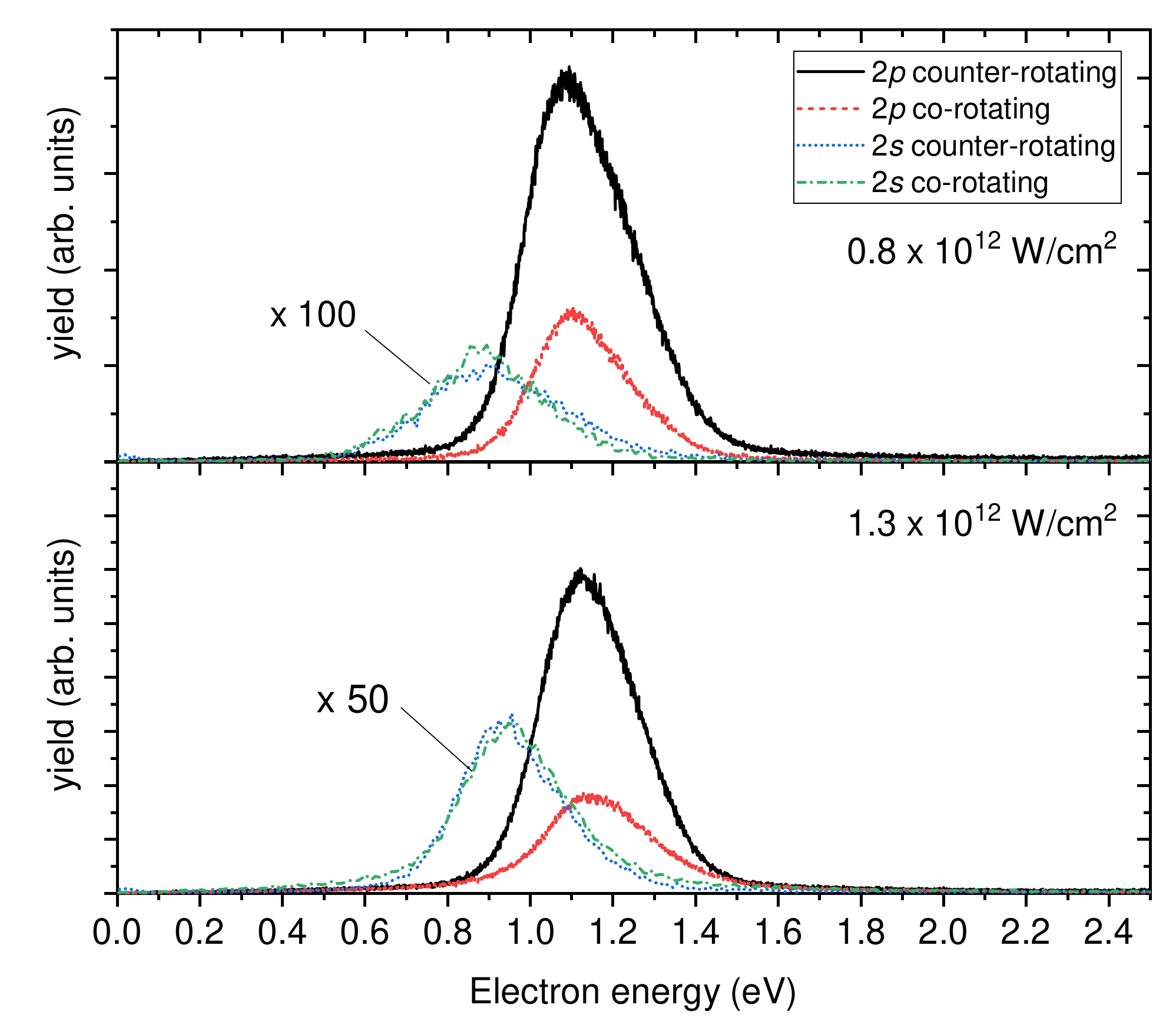}
\caption{Photoelectron energy distributions for 2$s$ and 2$p$ ionization by a 
laser field with a center wavelength of 795\,nm and peak intensities of 0.8$\times 10^{12}$\,W/cm$^2$ (top) 
and 1.3$\times10^{12}$\,W/cm$^2$ (bottom). \label{fig:energies}}
\end{figure}

Figure~\ref{fig:energies} depicts electron energy distributions for a 
laser wavelength of 795\,nm with intensities of 0.8$\times 10^{12}$\,W/cm$^2$ and 1.3$\times 10^{12}$\,W/cm$^2$. 
For 2$s$ ionization, the spectra exhibit a single peak around 0.9\,eV. 
For 2$p$ ionization, the peak is at slightly higher energies around 1.1\,eV. These values 
agree with the expectations for four- and three-photon ionization of the two respective 
initial states. In our earlier experiment studying the ionization of the same atomic 
system at a shorter wavelength close to the 2$s$-2$p$ resonance \cite{Silva2021}, 
intensity- and helicity-dependent photo\-electron energy shifts of up to 40\,\% were 
observed and explained by the Autler-Townes effect, which can be very strong in the 
vicinity of resonances. In the present case, such energy shifts are substantially smaller 
due to the much larger detuning of the laser center wavelength 
with respect to the resonance ($\Delta \lambda \approx 20$\,nm as compared to 5\,nm in 
the previous study) and the lower spectral density of the radiation (cf.\ Fig.~\ref{fig:Spectrum}).

%the photoelectron energy spectra shown in Fig.~\ref{fig:energies}
%are cross-normalized for each individual laser intensity. As described above, cross sections for 
%the ionization of the 2$s$ and the 2$p$ states for a given femtosecond laser
%handedness are obtained by switching the cooling laser periodically on and off.  
%The relative ionization rate for the two initial states is calculated exploiting the  
%information on the fraction of target atoms being in the excited state. The cross normalization between the two different
%laser field helicities is achieved, because the 2$s$ ionization cross sections are independent
%of the photon handedness and identical for left and right-handed polarization, 
%which can be seen by the closely matching blue and green curves in the figure.
%

The energy spectra shown in Fig.~\ref{fig:energies} are cross-normalized for the two laser field helicities exploiting the fact that the
ionization cross sections for the spherically symmetric 2$s$ initial state should be identical for both handednesses of the light.
As can be seen from the figure, the ionization of the 2$p$ state is strongly 
favored for the counter-rotating geometry. Using Eq.~(\ref{eq:Cd}), this helicity 
dependence can be quantified. Negative $CD$ values of about $-0.49\pm 0.1$ and $-0.54\pm 0.1$,
respectively, are obtained for both intensities.
The error provided here is an estimate considering the statistical error 
as well as the uncertainty in the excitation ratio of the atoms in the AOT, 
which is about $25\pm3$\,\%. Additional systematic uncertainties arise due 
to the imperfect polarization of both the laser radiation and the target atoms, 
as well as small drifts in the wavelength and intensity of the laser 
pulses during the measurements.

\begin{figure}
\centering
\includegraphics[width=1\linewidth]{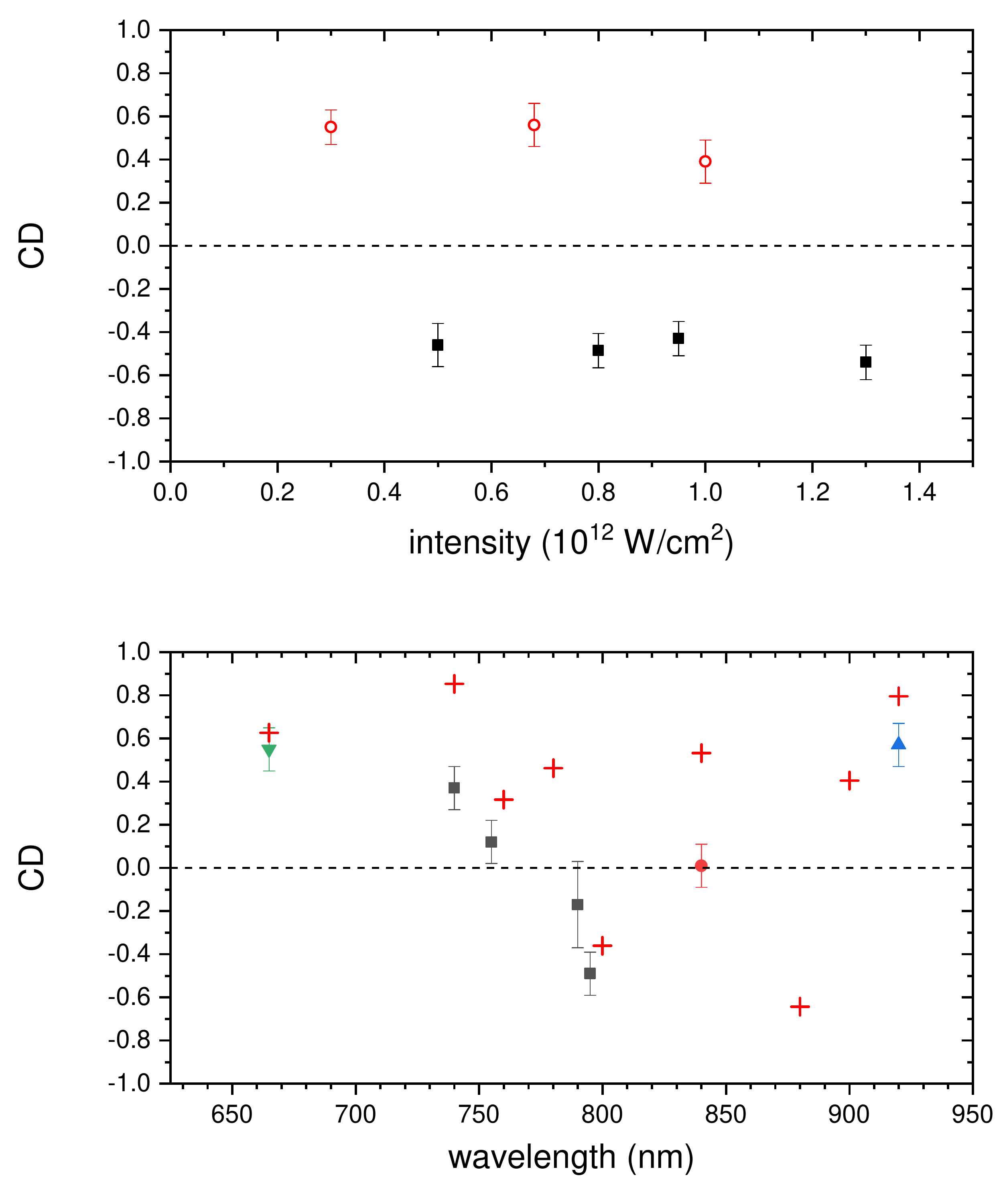}
\caption{Top: $CD$ values as a function of laser intensity for a center 
wavelength of 795\,nm (black solid squares) and 665\,nm (red open circles). 
Bottom: $CD$ values as a function of the center wavelength. The experimental 
data were measured for peak intensities of 0.3$\times 10^{12}$\,W/cm$^2$ (green and blue triangles), 0.4$\times 10^{12}$\,W/cm$^2$ (red circle), 
and 0.5$\times10^{12}$\,W/cm$^2$ (solid squares). The laser spectrum at 920\,nm was 
substantially broader ($\pm 70$\,nm) than in the other cases. The calculated $CD$ values 
(red crosses) are for 35\,fs (FWHM) Fourier-limited laser pulses with a peak intensity of 
0.3$\times 10^{12}$\,W/cm$^2$. \label{fig:CD}}
\end{figure} 

The intensity dependence of the $CD$ values was investigated in a series of 
measurements between 0.5$\times 10^{12}$\,W/cm$^2$ and 1.3$\times 10^{12}$\,W/cm$^2$ at a wavelength of $\lambda=795$~nm.
The obtained data are plotted in Fig.~\ref{fig:CD} (left) along with the values measured in 
our earlier experiment at $\lambda=665$\,nm, which are (partly) published 
in \cite{Silva2021, Silva2020}. For both wavelengths, the intensity dependence is fairly weak. 
It should be noted that it is not possible to measure the cross sections at a single 
intensity in our experiment due to the spatial extension of the reaction volume 
and the intensity distribution around the focal point of the laser beam. The intensities 
given in the graphs correspond to their highest values while the measured data  
represent averages over the intensity distribution in the focal volume \cite{Schuricke2011}. Notably, 
the $CD$ values have opposite sign for the two wavelengths. As mentioned above, 
the difference can be explained by the 2$p$-3$s$ resonance that enhances the 
ionization rate for the counter-rotating case around 815\,nm.

The above observation suggests that, in the intensity and wavelength regime considered, the dependence of the $CD$ on the 
wavelength of the incoming field is much stronger than its dependence on intensity. 
 To obtain further insight, we varied the wavelength of the laser field 
 between 740\,nm and 795\,nm at a constant intensity of about 5$\times 10^{11}$\,W/cm$^2$ 
and measured the corresponding $CD$ values. In this region, the $CD$ monotonically drops 
from positive to negative values (see Fig.~\ref{fig:CD}). 
 When further measurements at higher and lower wavelengths with slightly different spectral 
 widths and intensities were performed, a local minimum of the $CD$
 was observed at the 2$p$-3$s$ resonance.

The measured $CD$ values were compared with the results from the TDSE calculations.
We used 35\,fs FWHM (corresponding to a bandwidth of about 60\,meV) Fourier-limited pulses 
with Gaussian envelopes and a peak intensity of 0.3$\times10^{12}$\,W/cm$^2$. 
We observe an overall good agreement between the measurements and the calculations (see Fig.~\ref{fig:CD}). 
Note that the negative $CD$ value around 800\,nm is well reproduced theoretically. 
The calculation even predicts a second minimum with a 
negative $CD$ value near 880\,nm.  However, no experimental data are available at this wavelength.

To obtain a better understanding of the alternating behavior of the $CD$ values between 
800\,nm and 880\,nm, we monitored the population of the (undressed) 
bound states (with $n\leq 8$) in time. Even 
though it is not possible to unambiguously trace the population transfer from one specific 
state to another, this method still enables us to identify resonances that are relevant 
for ionization enhancement.

\begin{figure}
\centering
\includegraphics[width=1\linewidth]{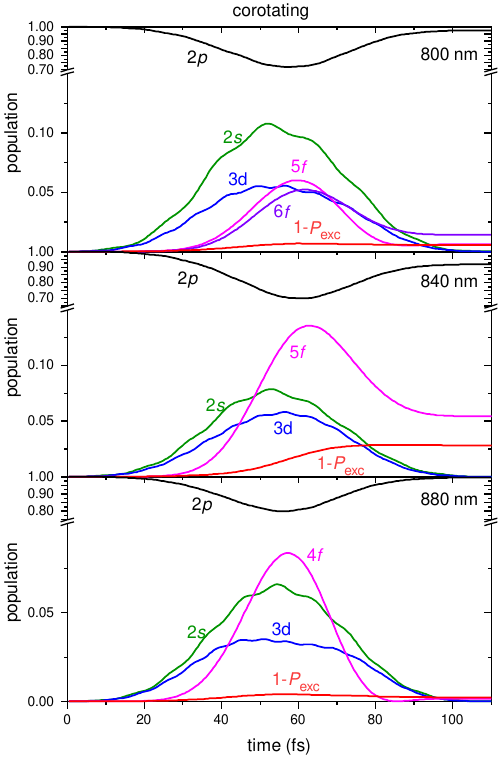}
\caption{Theoretical time-dependent atomic populations during a co-rotating pulse with 
peak intensities of 0.3$\times 10^{12}$\,W/cm$^2$ and center wavelengths of 800\,nm (top), 840\,nm (middle), and 880\,nm (bottom). 
The line denoted with $(1-P_\mathrm{exc})$ mainly corresponds to the population of the continuum states, although 
it contains a small contribution from states with $n>8$. \label{fig:popco}}
\end{figure} 

For the co-rotating field, the time-dependent populations are shown in Fig.~\ref{fig:popco} for 800\,nm, 
840\,nm, and 880\,nm light. The dressing of the initial $2p$ state by the 
weak laser field occurs mainly via one-photon coupling to the $2s$ and $3d$ states,
whose populations are seen to adiabatically follow the square of the pulse envelope,
 leading to the well-known quadratic AC stark shift.
It is noteworthy that these states mix strongly, although there is a large frequency detuning to the  $2s$-$2p$ and $2p$-$3d$
transitions at 671\,nm and 610\,nm, respectively. On the other hand, resonant transitions to the excited $5f$ and $6f$ Ryd\-berg states 
participate very effectively in the ionization enhancement.
These states, coupled to the $2p$ state by two-photon transitions,
undergo Rabi-like oscillations \cite{Gentile1989,Linskens1996,Fushitani2015}, whose frequency depends on the strength of the coupling,
leading to non\-negligible populations at the end of the pulse. The ionization enhancement is strongest at 
840\,nm due to the coupling between the 2$p$ and 5$f$ states, 
whose resonance wavelength is 826\,nm. In this case, the total 
ionization probability is more than five times higher than 
for the other two wavelengths, and a significant fraction of 
the population (more than 5\,\%) remains in the excited 5$f$ state after the pulse.
 At 880~nm, the $4f$ state is seen to mostly contribute to a quartic AC stark shift via 
two-photon coupling.

\begin{figure}
\centering
\includegraphics[width=1\linewidth]{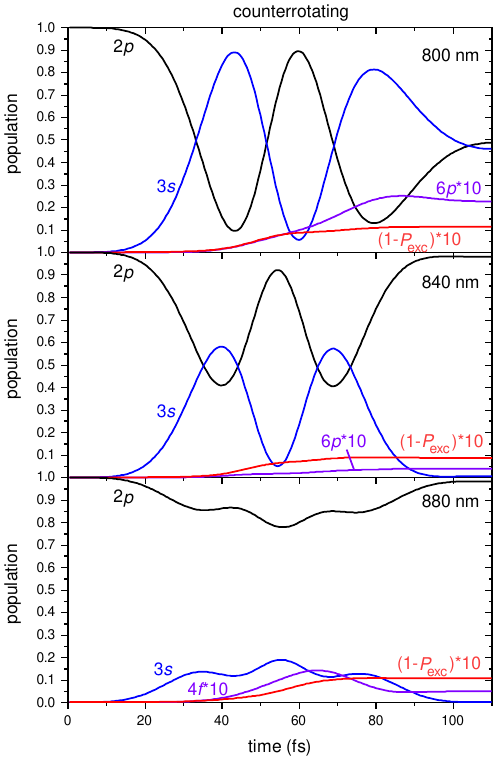}
\caption{Same as Fig.~\ref{fig:popco}, but for counter-rotating geometry. \label{fig:popcounter}}
\end{figure} 

The population distributions for the counter-rotating case are shown 
in Fig.~\ref{fig:popcounter}. As expected, prominent Rabi oscillations and a
large population transfer (up to  90\,\%) between the initial 2$p$ 
and the excited 3$s$ state appear at 800\,nm due to the proximity to the resonance.
 From the 3$s$ state, population can be transferred further up to the continuum using the 6$p$
 state as a stepping stone (see Fig.~\ref{fig:IonizationScheme}), resulting in an ionization probability well above 1\,\%. 
At 840\,nm, the overall picture remains similar with a weaker coupling between 
the 2$p$ and 3$s$ states leading to a slightly reduced ionization probability. 
Even further off resonance at 880\,nm, the amplitudes of the Rabi oscillations are strongly reduced 
and the 3s population follows the pulse envelope more adiabatically. 
It is surprising that the ionization rate increases again for this
wavelength.  However, there are two features of the population 
distributions that might explain the overall high ionization rate: First, 
the 3$s$ population is consistently above 10\,\% during the pulse and even approaches 20\,\% 
when the field intensity reaches its maximum. Second,
two-photon excitation of the 4$f$ state is observed as well, which might 
contribute to the large ionization probability at this wavelength. Overall, 
the variation of the ionization probability is much smaller for the counter-rotating 
geometry than for the co-rotating one. Therefore, we attribute the alternating behavior 
of the $CD$ mainly to the large fluctuations of the ionization probability for the co-rotating field.

This theoretical analysis shows that the ionization rate depends strongly
on the coherent population transfer between the initial and intermediate states. 
Therefore, it is in principle possible to induce sharp variations of the $CD$ through only 
small variations of the wavelength. Because the populations of the states undergo Rabi oscillations, the $CD$ is also affected 
by the intensity, duration, and wavelength of the pulse. Since the cross sections 
are not measured for a single peak intensity but for an intensity range spanning the laser focus,
along with other experimental uncertainties, our measured cross sections and $CD$ values represent an average that results
in a smoother variation of the $CD$ than in theory. These effects make the identification of intensity-related variations 
almost impossible, or at least very challenging, from an experimental perspective. 
Due to the remaining uncertainties and fluctuations in several of the experimental parameters, we did not perform additional calculations as they would most likely not shed more light on the interpretation of the experimental data.

It should be noted that there are also significant 
discrepancies between the measured and calculated $CD$ values at wavelengths 
of 740\,nm, 780\,nm, and 840\,nm. All these wavelengths are close 
to two-photon resonances between the 2$p$ and excited $np$ and $nf$ states, 
which lie around 826-829\,nm for $n=5$, 780\,nm for $n=6$, and 744\,nm for $n=8$. 
Although we did not perform a detailed investigation of the population distributions for these wavelengths, 
we expect that two-photon Rabi oscillations with periods close to the pulse duration 
are also strongly affecting the efficiency of resonance enhancement.

\section{Summary and Conclusions}

In this work, we addressed the question how atomic resonances affect the helicity dependence
of the multi-photon ionization of polarized atomic targets in circularly polarized femto\-second
laser fields. Specifically, we studied circular dichroism in two- and three-photon ionization of 
excited and polarized lithium atoms initially in the 2$p\,$($m_\ell=+1$) state for wavelengths
ranging from 665\,nm to 920\,nm. Experimentally, we find that the ionization probability is larger
for most wavelengths if the electric field vector and the initial electronic current density are 
co-rotating. However, there is a distinct minimum in the $CD$ value at a wavelength near 800\,nm, 
where the counter-rotating geometry is favored. This feature is qualitatively explained 
by the 2$p$--3$s$ resonance, which is only driven for the counter-rotating case, thus resulting
in a helicity-selective resonance enhancement.

The comparison of the experimental data to our theoretical model, which is based on the numerical
 solution of the time-dependent Schr{\"o}dinger equation, yields good qualitative agreement, although the calculation shows a 
larger scattering of the $CD$ values and even a second minimum around 880\,nm. The theoretical
approach allows to extract time-dependent population distributions of the atomic states during 
and after the femtosecond laser pulse, which helps to shed light on the under\-lying dynamics 
and the involved intermediate states. This analysis shows Rabi oscillations between states 
coupled by one- and two-photon transitions. The populated excited states can result in resonance-enhancement and act as stepping stones
for the electrons while they are promoted to the continuum.
Because Rabi frequencies and time-dependent excited state populations depend sensitively on the details of the pulse,
such as its intensity, duration, and detuning off a resonance, the total ionization probabilities and, therefore, 
the $CD$ values are expected to be strongly influenced  by these parameters as well. In our experiment, the peak intensity of the laser pulse is difficult to control
due to the spatial intensity profile of the laser focus in the reaction region. As a result, the experimental $CD$ values represent
a weighted average for different intensities.

The present work does not only show the importance of the atomic structure on the ionization probability and the circular dichroism, but it also exposes
the significance of details of the pulse parameters on the ionization dynamics.

\medskip

\section*{Acknowledgments}
The experimental material presented here is based
upon work supported by the National Science Foundation
under Grant \hbox{No.~PHY-1554776}.
The theoretical part of this work was funded by the NSF under
grants \hbox{No.~PHY-2012078} (T.M.\ and N.D.) and \hbox{PHY-1803844} (K.B.), and by the XSEDE supercomputer allocation No.~PHY-090031. 
The calculations were carried out on Comet at the San Diego Supercomputer Center and \hbox{Bridges-2} (via a trial allocation)
at the Pittsburgh Super\-computing Center.

\end{document}